\documentclass{JHEP3}

\usepackage{amsmath}
\usepackage{amsfonts}
\allowdisplaybreaks

\newcommand{\PSU}{\mathop{\rm PSU}\nolimits}
\newcommand{\SO}{\mathop{\rm SO}\nolimits}
\newcommand{\GL}{\mathop{\rm GL}\nolimits}

\title{The algebra of transition matrices for the $AdS_{5} \times
  S^{5}$ superstring}

\author{Ashok Das, Arsen~Melikyan and Matsuo Sato\\
  Department of Physics and Astronomy, University of Rochester\\
  Rochester, NY 14627-0171, U.S.A.\\
  E-mail: \email{das@pas.rochester.edu},
          \email{arsen@pas.rochester.edu},
	  \email{sato@pas.rochester.edu}}

\author{Jnanadeva Maharana\\
  Institute of Physics\\
  Bhubaneswar-751006, India\\
  E-mail: \email{jnanadeva.maharana@cern.ch}}

\abstract{We consider integrability properties of the superstring on
  $AdS_{5}\times S^{5}$ background and construct a new one parameter
  family of currents which satisfies the vanishing curvature
  condition. We present the hamiltonian analysis for the sigma model
  action and determine the Poisson algebra of the transition
  matrices. We reveal the generalization of the $\mathbb{Z}_{4}$
  automorphism analogous to the sigma models defined on a symmetric
  space coset. A possible regularization scheme for the ambiguities
  present, which respects the generalized automorphism, is also
  discussed.}

\received{November 30, 2004}
\accepted{December 21, 2004}

\begin{document}

\section{Introduction}\label{section1}

The $AdS/CFT$ correspondence~\cite{mald}--\cite{rev} has proved to be
very crucial in establishing correspondence between superstring and
supersymmetric gauge theories. The $AdS/CFT$ correspondence becomes
tractable essentially in two different situations.  Namely, when the
't Hooft coupling ($g_{YM}^{2}N$) is small, one can adopt the large
$N$ technique to carry out computations in the superYang-Mills (SYM)
sector~\cite{buch}--\cite{agarwal}. On the other hand, for the large
values of 't Hooft coupling, the duality argument can be invoked to
facilitate computations in the string theory perturbation (inverse
tension) frame work~\cite{sato}--\cite{add4}.  For values of the 't
Hooft coupling which are neither small nor large, there is no obvious
simplification.  In a recent paper, Bena, Polchinski and
Roiban~\cite{PBR} studied the integrability properties of the
superstring propagating on the $AdS_{5}\times S^{5}$ background with a
goal of exploring the dynamics of the string theory for such an
intermediate coupling when the string theory is not necessarily in the
perturbative regime.

The integrability properties of SYM have been subject of considerable
interest in recent years. The conformal group in four dimensions is
identified to be $\SO(4,2)$ and dilatation is one of its
generators. It is well known in various computations to establish
$AdS/CFT$ correspondence, the Yang-Mills theory lives on
$\mathbb{R}\times S^{3}$ and the dilatation operator $D$ is the
hamiltonian in the radial quantization scheme~\cite{fubini}. Minahan
and Zarembo~\cite{mz}--\cite{beis5} computed anomalous dimensions of a 
class
composite operators of ${\cal N}=4$ SYM in the large $N$ limit at the
one loop level. They show that the anomalous dimension operator, in
the radial quantization, is related to a spin chain quantum
hamiltonian which is known to be integrable.  There is also
correspondence between semiclassical string states and composite
operators of SYM~\cite{nb}. It has been argued by Dolan, Nappi and
Witten~\cite{dnw} that the integrability property of SYM unraveled in
this context is intimately connected with the yangian symmetry (see
for a review~\cite{mackay}) associated with ${\cal N}=4$ SYM when one
sets the Yang-Mills coupling to zero. Furthermore, they conjecture
that the yangian symmetry discovered for the SYM is related to the
yangian symmetry which one expects to obtain due to the presence of
conserved nonlocal currents for the superstring when it propagates in
$ AdS_{5}\times S^{5}$ background. Therefore, the hidden symmetries
uncovered by Bena, Polchinski and Roiban~\cite{PBR} are connected with
the integrability properties of SYM discussed above.

If one considers a bosonic string in an $AdS_{n}$ space, the
worldsheet action may be identified with an $O(n)$ nonlinear $\sigma
$-model~\cite{msw} and one may adopt well known prescriptions to
construct a family of nonlocal conserved currents~\cite{chiral} for
such a case which is responsible for the classical integrability of
the model. Let us recall how the $AdS_{5}\times S^{5}$ geometry arises
in type IIB string theory. The $5$-dimensional extremal black hole
solution is obtained with appropriate choice of backgrounds which
solve the equations of motion. Of special significance, in this case,
is the fact that the five form $RR$ flux and the dilaton assume
constant values. Subsequently, the near horizon limit is taken which
leads to the geometry of $AdS_{5}\times S^{5}$. Therefore, the
worldsheet action for such a theory is to be constructed keeping in
mind the presence of constant $RR$ background.

In such a case, the standard technique of~\cite{chiral} becomes
inadequate. This is due to the fact that the $NSR$ formalism is
inapplicable to construct a suitable nonlinear $\sigma $-model action
on the worldsheet. Consequently, it is the Green-Schwarz formalism
that is more appropriate in this context where the theory can be
described as a nonlinear sigma model with a Wess-Zumino-Witten
term. In this case, the basic field variables of the action
parametrize the coset
\begin{equation}
\frac{{\PSU(2,2|4)}}{{\SO(4,1)\times \SO(5)}}\,,  \label{cosett}
\end{equation}
as has been shown by several authors~\cite{MT}--\cite{rs}. It is worth
emphasizing here that the Green-Schwarz action for the superstring on
$AdS_{5}\times S^{5}$ is not strictly a coset $\sigma$-model (unlike
the bosonic theories) due the presence of the fermionic Wess-Zumino
term as well as the local $\kappa$ symmetry. The properties of the
coset and its grading structure played a key role in construction of
the nonlocal currents~\cite{PBR}. Recently, Hatsuda and
Yoshida~\cite{KEK} have taken into account the presence of the
Wess-Zumino term and constructed explicitly the nonlocal charges as
well as the yangian algebra associated with the system. Their work
involves a euclideanized supergroup $\GL(4|4)$ and the sigma model is
defined on a superspace.

In studying the integrability properties of a sigma model, the basic
object is a one parameter family of currents which satisfies a zero
curvature condition. There is, however, an arbitrariness in the choice
of this current. In construction of objects such as the monodromy
matrix, it is the current invariant under a generalized inner
automorphism of the symmetry group that plays a crucial
role~\cite{nicolai}. This symmetry is, however, lacking in the one
parameter family of currents constructed in~\cite{PBR}. In this paper,
therefore, we have tried to study the integrability properties of the
superstring on the $AdS_{5}\times S^{5}$ background keeping the
conventional automorphisms manifest. We construct a one parameter
family of currents which manifestly is invariant under the inner
automorphism of the graded group $\mathbb{Z}_{4}$. In spite of the
fact that the sigma model in this case is not a genuine coset space
model, we find a generalization of the inner automorphism to
$\mathbb{Z}_{4}^{\infty}$ which is relevant in the construction of the
monodromy matrix. We systematically construct the nonlocal charges
from this current. Using the conventional sigma model action (in terms
of currents) with a Wess-Zumino-Witten term in the coset
space~\cite{BBHZZ}, we carry out the hamiltonian analysis of the
system and determine the basic Poisson brackets of the current (this
generates the yangian algebra) which is essential in the construction
of the algebra of the transition matrix and which has a closed
form. We point out the difficulties that arise due to the presence of
the $\kappa$ symmetry in the action. In addition, we clarify some
subtleties associated with the Virasoro constraints of the theory in
this case.

The paper is organized as follows. In section~\ref{section2} we
recapitulate the basic properties of the type IIB Green-Schwarz
superstring action on the $AdS_{5}\times S^{5}$ background. In
section~\ref{section3} we present essential properties of the
superalgebra $\PSU(2,2|4)$. In section~\ref{section4} we introduce a
one parameter family of currents which satisfies the vanishing
curvature condition and has a form that reveals the special ${\mathbb
  Z}_{4}^{\infty }$ automorphism. In section~\ref{section5} we
construct the hierarchy of conserved nonlocal currents. In the final
section~\ref{section6} we perform the hamiltonian analysis and
calculate the Poisson bracket of the flat currents. We discuss various
aspects of the results as well as future directions for this analysis
and conclude with a brief summary in section~\ref{section7}.

\section{Superstring on $AdS_{5}\times S^{5}$}\label{section2}

We summarize here some of the basic properties of the type IIB
Green-Schwarz superstring action on the $AdS_{5}\times S^{5}$
background~\cite{MT, krr, kr2, KalTs}. The superstring can be defined
as a non-linear sigma model on the coset superspace
\begin{equation}
\frac{G}{H}=\frac{\PSU(2,2|4)}{\SO(4,1)\times \SO(5)}\,.  \label{coset}
\end{equation}
The classical action has the Wess-Zumino-Witten form
\begin{equation}
S=-\frac{1}{2}\underset{\partial M^{3}}{\int}d^{2}\sigma
\sqrt{-g}g^{ij}\left( L_{i}^{\hat{a}}L_{j}^{\hat{a}}\right)
+i\underset{M^{3}}{\int} s^{IJ}\left(
L^{\hat{a}}\wedge\overline{L}^{I}\gamma^{\hat{a}}\wedge L^{J}\right),
\label{MT}
\end{equation}
where $g^{ij}, i,j=0,1$ represents the worldsheet metric, $s^{IJ}={\rm
  diag} (1,-1), I,J=( 1,2); \hat{a}=(a,a^{\prime })$ with $a=(
0,\dots,4)$ and $a^{\prime}=( 5,\dots,9)$ corresponding to tangent
space indices for $AdS_{5}$ and $S^{5}$ respectively. We use the
convention that repeated indices are summed. The supervielbeins
$L^{\hat{a}}$ and $L^{I}$ are defined as
\begin{eqnarray}
L^{I} & =&\left( \left( \frac{\sinh\mathcal{M}}{\mathcal{M}}\right)
D\theta\right) ^{I}\,,  \nonumber\\
L^{\hat{a}} & =&e_{\hat{\mu}}^{\hat{a}}(x)\mathrm{d}x^{\hat{\mu}}-
i\overline{\theta} \gamma^{ \hat{a}}\left(\left( \frac{\sinh
  \mathcal{M}/2}{\mathcal{M}/2}\right)^{2} D\theta\right),
\label{supervielbeins} 
\end{eqnarray}
where
\begin{equation}
\left( \mathcal{M}^{2}\right) ^{IJ}  = \epsilon^{IK}\left( -\gamma
^{a}\theta^{K}\bar{\theta}^{J}\gamma^{a}+\gamma^{a^{\prime}}\theta^{K}
\bar{\theta}^{J}\gamma^{a^{\prime}}\right)  +
\frac{1}{2}\epsilon^{KJ}\left( \gamma^{ab}\theta^{I}\bar{\theta}
^{K}\gamma^{ab}-\gamma^{a^{\prime}b^{\prime}}\theta^{I}\bar{\theta}^{K}
\gamma^{a^{\prime}b^{\prime}}\right).  
\label{M}
\end{equation}
Here $( x^{\hat{\mu}},\theta^{I}) $ denote the bosonic and fermionic
string coordinates in the target space, $(
e^{\hat{a}},\omega^{\hat{a}\hat{b}}) $ are the bosonic vielbein and
the spin connection respectively and the covariant differential is
given by
\begin{equation}
\left( D\theta\right)
^{I}=\left[\delta^{IJ}\left(\mathrm{d}+\frac{1}{4}\
  \omega^{\hat{a}\hat{b}} \gamma_{\hat{a}\hat{b}}\right)
  -\frac{i}{2}\epsilon^{IJ}\ e^{\hat{a}}\gamma_{\hat{a}} \right]
\theta^{J}\,.  \label{covdif}
\end{equation}

The equations of motion following from the action (\ref{MT}) take the
forms
\begin{eqnarray}
\sqrt{-g}g^{ij}\left( \nabla_{i}L_{j}^{a}+L_{i}^{ab}L_{j}
^{b}\right) +i\epsilon^{ij}s^{IJ}\bar{L}_{i}^{I}\gamma^{a}L_{j}^{J} & =&0\,,
\label{eqofm1} \\[-2pt]
\sqrt{-g}g^{ij}\left( \nabla_{i}L_{j}^{a^{\prime}}+L_{i}^{a^{\prime
}b^{\prime}}L_{j}^{b^{\prime}}\right) -\epsilon^{ij}s^{IJ}\bar{L}_{i}
^{I}\gamma^{a^{\prime}}L_{j}^{J} & =&0\,,  \label{eqofm2} \\[-2pt]
\left( \gamma^{a}L_{i}^{a}+i\gamma^{a^{\prime}}L_{i}^{a^{\prime}}\right)
\left( \sqrt{-g}g^{ij}\delta^{IJ}-\epsilon^{ij}s^{IJ}\right) L_{j}^{J} & =&0\,,
\label{eqofm3}
\end{eqnarray}
with $\nabla_{i}$ representing the covariant derivative on the
worldsheet. We will use the conformal gauge
$\sqrt{-g}g^{ij}=\eta^{ij}$ in which case the equations of
motion~(\ref{eqofm1})--(\ref{eqofm3}) simplify, but should be
complemented with the Virasoro constraints
\begin{equation}
L_{i}^{\hat{a}}L_{j}^{\hat{a}} = \frac{1}{2}\ \eta _{ij}\eta^{kl}\
L_{k}^{\hat{a}}L_{l}^{\hat{a}}\,.  \label{virasoroconst}
\end{equation}
To consider the integrability properties of the sigma model we will
need some properties of the superalgebra $\PSU(2,2|4)$ which we review
in the next section.

\section{Properties of $\PSU(2,2|4)$}\label{section3}

In this section we discuss some of the essential properties of the
superalgebra $\PSU (2,2|4)$ \cite{BBHZZ}
and~\cite{kac3}--\cite{corn}. Since we are interested in a
supersymmetric field theory, we assume that the algebra is defined on
a Grassmann space, $\PSU (2,2|4;\mathbb{C}B_{L})$. We represent an
element of this superalgebra by an even supermatrix of the form
\begin{equation}
G =\left( 
\begin{array}{cc}
A & X \\ 
Y & B
\end{array}
\right),  \label{mat1}
\end{equation}
where $A$ and $B$ are matrices with Grassmann even functions while $X$
and $Y$ are those with Grassmann odd functions, each representing a
$4\times 4$ matrix. (An odd supermatrix, on the other hand, has the
same form, with $A$ and $B$ consisting of Grassmann odd functions
while $X$and $Y$ consisting of Grassmann even functions.)

An element $G$ (see~\ref{mat1}) of the superalgebra $\PSU (2,2|4;
\mathbb{C} B_{L})$ is given by a 8 $\times$ 8 matrix, satisfying 
\begin{eqnarray}
GK+KG^{\ddagger}&=&0\,,  \label{antihermite} \\
\mbox{tr}A=\mbox{tr}B&=&0\,,  \label{mat3}
\end{eqnarray}
where {\scriptsize{$K=\left( 
\begin{array}{cc}
\Sigma & 0 \\ 
0 & I_{4}
\end{array}
\right)$}} and $\Sigma= \sigma_{3}\otimes I_{2}$ with $I_{2},I_{4}$
    representing the identity matrix in $2$ and $4$ dimensions
    respectively. The $\ddagger$ is defined by
\begin{equation}
G^{\ddagger}=G^{{\rm T} \sharp}\,,  \label{mat4}
\end{equation}
where ${\rm T}$ denotes transposition and $\sharp$ is a generalization
of complex conjugation which acts on the functions $c$ of the matrices
as
\begin{equation}
c^{\sharp}=\left\{ 
\begin{array}{cl}
c^{\ast} & \mbox{(for $c$ Grassmann even)} \\ 
-ic^{\ast} & \mbox{(for $c$ Grassmann odd)}
\end{array}
\right. .  \label{mat5}
\end{equation}
The condition (\ref{antihermite}) can be written explicitly as
\begin{equation}
\Sigma A^{\dagger}+A\Sigma=0\,,\qquad B^{\dagger}+B=0\,,\qquad X-i\Sigma
Y^{\dagger }=0\,.  \label{mat5a}
\end{equation}

The essential feature of the superalgebra $\PSU (2,2|4)$ is that it
admits a $\mathbb{Z}_{4}$ automorphism such that the condition
$\mathbb{Z}_{4} (H) = H$ determines the maximal subgroup to be
$\SO(4,1)\times \SO(5)$ which leads to the definition of the coset for
the sigma model. (This is the generalization of the $\mathbb{Z}_{2}$
automorphism of bosonic sigma models to the supersymmetric case.) The
$\mathbb{Z}_{4}$ automorphism $\Omega$ takes an element of $\PSU
(2,2|4)$ to another, $G \rightarrow \Omega (G)$, such that
\begin{equation}
\Omega (G) =\left( 
\begin{array}{cc}
JA^{\rm T}J & -JY^{\rm T}J \\ 
JX^{\rm T}J & JB^{\rm T}J
\end{array}
\right),  \label{mat6}
\end{equation}
where {\scriptsize{$J=\left( 
\begin{array}{cc}
0 & -1 \\ 1 & 0
\end{array}
\right)$}}. It follows now that $\Omega^{4} (G) = G$.

Since $\Omega^{4}=1$, the eigenvalues of $\Omega$ are $i^{p}$ with $p
=0,1,2,3$. Therefore, we can decompose the superalgebra as
\begin{equation}
G= \mathcal{H}_{0}\oplus \mathcal{H}_{1}\oplus \mathcal{H}_{2}\oplus
\mathcal{H}_{3}\,, \label{dec}
\end{equation}
where $\mathcal{H}_{p}$ denotes the eigenspace of $\Omega$ such that
if $H_{p}\in \mathcal{H}_{p}$, then
\begin{equation}
\Omega(H_{p})= i^{p}H_{p}\,.  \label{eig}
\end{equation}
We have already noted that $\Omega (\mathcal{H}_{0}) =
\mathcal{H}_{0}$ determines $\mathcal{H}_{0} = \SO(4,1)\times
\SO(5)$. $\mathcal{H}_{2}$ represents the remaining bosonic generators
of the superalgebra while $\mathcal{H}_{1},\mathcal{H}_{3}$ consist of
the fermionic generators of the algebra. (In a bosonic sigma model,
$\mathcal{H}_{0},\mathcal{H}_{2}$ are represented respectively as
$Q,P$.)  The automorphism also implies that
\begin{equation}
\lbrack H_{p},H_{q}]\in \mathcal{H}_{p+q\pmod{4}}.  \label{mat8}
\end{equation}

The space $\mathcal{H}_{p}$ is spanned by the generators $(t_{p})_{A}$
of the superalgebra so that we can explicitly write 
\begin{eqnarray}
G & = & (H_{p})^{A}(t_{p})_{A}\nonumber\\
 & = & (H_{0})^{m}(t_{0})_{m}+(H_{1})^{\alpha_{1}}
(t_{1})_{\alpha_{1}}+(H_{2})^{a}(t_{2})_{a}+(H_{3})^{\alpha_{2}}(t_{3})_{
\alpha_{2}}\,,\label{mat9}
\end{eqnarray}
where $A=(m,\alpha_{1},a,\alpha_{2})$ take values over all the
generators of the superalgebra, $(H_{0})^{m}$ and $(H_{2})^{a}$ are
Grassmann even functions, while $ (H_{1})^{\alpha1}$ and
$(H_{3})^{\alpha2}$ are Grassmann odd functions. The generators
satisfy the graded algebra $\PSU(2,2|4)$,
\begin{equation}
\lbrack(t_{p})_{A},(t_{q})_{B}]=f_{AB}^{\quad C}(t_{p+q})_{C}\,,  \label{mat10}
\end{equation}
where $p+q$ on the right hand side is to be understood modulo $4$.

The Killing form (or the bilinear form) $\langle H_{p},H_{q}\rangle $
is also $\mathbb{Z}_{4}$ invariant so that
\begin{equation}
\langle \Omega\left( H_{p}\right) ,\Omega\left( H_{q}\right)\rangle =
\langle H_{p},H_{q}\rangle\,.
\end{equation}
This implies that
\begin{equation}
i^{(p+q)} \langle H_{p},H_{q}\rangle = \langle H_{p},H_{q}\rangle\,,
\label{killing}
\end{equation}
which leads to
\begin{equation}
\langle H_{p},H_{q}\rangle =0\quad {\rm unless}\quad p+q = 0\ \pmod{4}.  \label{form}
\end{equation}
Since the supertrace of a supermatrix $M$ is defined as  
\begin{equation}
\mbox{str}(M) = \left\{ 
\begin{array}{cl}
{\rm tr} A-{\rm tr} B & \mbox{(if $M$ is an even supermatrix)} \\ 
{\rm tr} A+{\rm tr} B & \mbox{(if $M$ is an odd supermatrix)}
\end{array}
\right. ,  \label{mat11}
\end{equation}
and the metric of the algebra is defined as $G_{AB} = {\rm
  str}((t_{p})_{A}(t_{q})_{B})$, the above relation also implies that
only the components $G_{mn},G_{ab}
,G_{\alpha_{1}\alpha_{2}}=-G_{\alpha_{2}\alpha_{1}}$ of the metric are
non-zero.  The structure constants possess the graded anti-symmetry
property
\begin{equation}
f_{AB}^{D} G_{DC} = - (-)^{|A| |B|}f_{BA}^{D} G_{DC} = - (-)^{|B|
  |C|}f_{AC}^{D} G_{DB}\,,
\end{equation}
where $|A|$ denotes the Garssmann parity of $A$, namely, $|A|$ is 0
when $A$ is $m$ or $a$, while $|A|$ is 1 when $A$ is $\alpha_{1}$ or
$\alpha_{2}$.

\section{The flat current}\label{section4}

Let us consider the map $g$ from the string worldsheet into the graded
group $\PSU(2,2|4)$. In this case, the current $1$ form
$J=-g^{-1}\mathrm{d}g$ belongs to the superalgebra and therefore can
be decomposed as
\begin{equation}
J=-g^{-1}\mathrm{d} g = H+P+Q^{1}+Q^{2}\,,  \label{curdecomp1}
\end{equation}
where, in terms of our earlier notation, we can identify
\begin{equation}
H=H_{0}\,,\qquad Q^{1}=H_{1}\,,\qquad P=H_{2}\,,\qquad Q^{2}=H_{3}\,.
\label{ident}
\end{equation}
\looseness=1 From the definition of the current in~(\ref{curdecomp1}), we see that
it satisfies a zero curvature condition
\begin{equation}
\mathrm{d} J - J \wedge J = 0\,.\label{zc}
\end{equation}
In terms of the components of the current (\ref{ident}), the equations
of motion can be written as~\cite{PBR}
\begin{eqnarray}
\mathrm{d}\ {}^{\ast}P & =&{}^{\ast}P \wedge H + H\wedge {}^{\ast}P +
\frac{1}{2}  \left( Q\wedge
Q^{\prime}+Q^{\prime}\wedge Q\right),  \nonumber\\
0 & =& P\wedge({}^{\ast}Q - Q^{\prime})+({}^{\ast}Q - Q^{\prime})\wedge
P\,, 
\nonumber\\
0 & =& P\wedge(Q - {}^{\ast}Q^{\prime})+(Q - {}^{\ast}Q^{\prime})\wedge
P\,,
\label{eqom1} 
\end{eqnarray}
where $^{\ast}$ denotes the Hodge star operation and we have defined
$Q\equiv Q^{1}+Q^{2},Q^{\prime}\equiv Q^{1}-Q^{2}$.

Let us next introduce a one parameter family of currents
$\hat{J}(t)\equiv-\hat{g} ^{-1}(t)\mathrm{d}\hat{g}(t)$ where $t$ is a
constant spectral parameter (here we are suppressing the dependence on
the worldsheet coordinates)
\begin{equation}
\hat{J}(t)=H+\frac{1+t^{2}}{1-t^{2}}\ P+\frac{2t}{1-t^{2}}\
    {}^{\ast}P+ \sqrt { \frac{1}{1-t^{2}}}\
    Q+\sqrt{\frac{t^{2}}{1-t^{2}}}\ Q^{\prime}\,,
\label{flatcurrent}
\end{equation}
such that $\hat{J} (t=0) = J$. It is easy to check that the vanishing
curvature condition for this new current
\begin{equation}
\mathrm{d}\hat{J}-\hat{J}\wedge\hat {J} = 0\,,
\end{equation} 
leads to all the equations of motion~(\ref{eqom1}) as well as the zero
curvature condition~(\ref{zc}).

It is worth comparing the form of this one parameter family of
currents~(\ref{flatcurrent}) with that for a (bosonic) two dimensional
sigma model obtained from the string effective action dimensionally
reduced from $D$-dimensions ~\cite{jmj}. In this case, the sigma model
coupled to gravity is defined on the symmetric space $\frac{G}{H}=$
$\frac{O(d,d)}{O(d)\times O(d)}$ where $d=D-2$ and the corresponding
one parameter family of currents has the form (conventionally in a
bosonic sigma model $\hat{g}$ is written as $\hat{V}$, but we are
using the same letter for ease of comparison)
\begin{equation}
\hat{J}(t)=\hat{g}^{-1}(t)\mathrm{d}\hat{g}(t) = H +
\frac{1+t^{2}}{1-t^{2}}\ P+\frac {2t }{1-t^{2}}\ {}^{\ast}P\,,
\label{bosoniccurrent}
\end{equation}
where the components of the currents belong to the appropriate
spaces. This has the same form as~(\ref{flatcurrent}) if we set the
fermionic generators to zero (the groups are, of course,
different). However, apart from the presence of the fermionic degrees
of freedom in the $AdS_{5}\times S^{5}$ theory, the essential
difference between the two theories lies in the fact that when we
dimensionally reduce the string effective action to two space-time
dimensions, it describes a sigma model coupled to gravity. As is well
known, in such a theory the spectral parameter assumes space-time
dependence for consistency.  Namely, the consistent zero curvature
description in this case requires that the spectral parameter
satisfies the condition
\begin{equation}
\partial_{\alpha}\rho=-\frac{1}{2}\varepsilon_{\alpha\beta}\partial^{\beta
}\left( e^{-\phi}\left( t+\frac{1}{t}\right) \right),  \label{spectcondition}
\end{equation}
where $\phi$ is the shifted dilaton and $\rho=e^{-\phi}$.  In
contrast, in the case of $AdS_{5}\times S^{5}$, the spectral parameter
is a global parameter.

There is, however, an important symmetry which the one parameter
family of currents in both the theories share. In the bosonic sigma
model coupled to gravity, the theory is invariant under a
generalization of the symmetric space automorphism (which is
$\mathbb{Z}_{2}$ as we have alluded to earlier)
\begin{equation}
\eta^{\infty}\left( \hat{g}(t)\right) =\left( \hat{g}^{-1}\left(\frac{1}
    {t} \right)\right) ^{\rm T}\,.  \label{auto1}
\end{equation}
This symmetry is essential in the construction of the monodromy matrix
$\mathcal{M}=\hat{g}(t)\hat{g}^{\rm T}(\frac{1}{t})$ \cite{nicolai,
  DMM} which encodes integrability properties and is related to the
transition matrix~\cite{bm, KS}.  In the case of the superstring on
the $AdS_{5}\times S^{5}$ background, the coset~(\ref{coset}) is not
exactly a symmetric space. We can, however, still define a
generalization of the $\mathbb{Z}_{4}$ automorphism of the
superalgebra to $\mathbb{Z} _{4}^{\infty}$ in a manner analogous
to~(\ref{auto1}). Explicitly, it is easy to check that the one
parameter family of currents is invariant under (we choose $\sqrt{-1}
= i$)
\begin{eqnarray}
t & \longrightarrow& \frac{1}{t}\,,  \nonumber\\
H & \longrightarrow& H\,,\qquad P\longrightarrow-P\,,  
\nonumber\\
Q & \longrightarrow& -i Q^{\prime},\quad Q^{\prime}\longrightarrow -i
Q\,. 
\label{symmet3} 
\end{eqnarray}
Consequently, this current is different from the one constructed
in~\cite{PBR} which does not have the necessary invariance property
under the inner automorphism. (There is an arbitrariness in the choice
of the one parameter family of flat currents and we have constructed
one that has the desired behavior under the inner automorphism of the
symmetry algebra which parallels the construction in conventional
sigma models.)  This symmetry will allow us to define the monodromy
matrix for the present case in a way similar to the symmetric space
bosonic sigma model~\cite{DMM} and is, consequently, quite
important. We note that one of the fermionic constraints
in~(\ref{eqom1}) is, in fact, a consequence of this
$\mathbb{Z}_{4}^{\infty}$ symmetry.

\section{Nonlocal currents}\label{section5}

The sigma model action which leads to the equations of motion
(\ref{eqom1}) has the form
\begin{equation}
S=\frac{1}{2}\int{\rm str}\left(P\wedge {}^{\ast}P - Q^{1}\wedge
{Q^{2}}\right),   \label{newaction}
\end{equation}
where $P, Q^{1}, Q^{2}$ are defined in~(\ref{curdecomp1}).  The first
term in the action~(\ref{newaction}) is similar to that in the
principal chiral model, while the second represents the WZW
term~\cite{BBHZZ}. This action is manifestly invariant under a local
(gauge) transformation $g\rightarrow gh$ where $h\in \SO(4,1)\times
\SO(5)$ is a local function, since under such a transformation
\begin{eqnarray}
H & \longrightarrow& h^{-1}H\ h - h^{-1}\mathrm{d}h\,,  
\nonumber\\
P & \longrightarrow& h^{-1} P\ h\,,
\nonumber\\
Q^{1,2} & \longrightarrow& h^{-1} Q^{1,2}\ h\,. 
\label{gaugetrans} 
\end{eqnarray}
The action is also invariant under a global transformation (left
multiplication) $g\rightarrow \omega g$ where $\omega\in \PSU(2,2|4)$
and the corresponding conserved Noether current for this left
translation can be obtained from the action~(\ref{newaction}) to be
\begin{equation}
j^{(0)} = p+ \frac{1}{2} {}^{\ast}q^{\prime},\quad \mathrm{d}\
{}^{\ast}j^{(0)} = 0\,.\label{noether}
\end{equation}
(The meaning of the superscript will become clear shortly, namely, it
is the zeroth order current in the infinite hierarchy of conserved
currents.)  Here we have adopted the notations of~\cite{PBR} for the
left and right invariant currents and the relation
\begin{equation}
x = g X g^{-1}\,,  \label{convrule}
\end{equation}
where lower and upper case objects denote left and right invariant
quantities respectively.

\pagebreak[3]

We can now determine the conserved non-local charges associated with
the system easily following~\cite{CZ} and~\cite{dasbnm, jm}. Let us
note that the covariant derivative
\begin{equation}
\hat{D}_{\mu} \equiv \partial_{\mu} - \hat{J}_{\mu} (t)\,,
\label{covd}
\end{equation}
satisfies the zero curvature condition
\begin{equation}
\lbrack \hat{D}_{\mu} , \hat{D}_{\nu}]=  0\,.  \label{zerocurv}
\end{equation}
As a result, the equation
\begin{equation}
\hat{D}_{\mu}\chi=\partial_{\mu}\chi-\hat{J}_{\mu}(t)\chi=0\,,
\label{int} 
\end{equation}
is integrable. It is convenient to rewrite~(\ref{int}) in the
following equivalent form
\begin{equation}
\varepsilon_{\mu\nu}\partial^{\nu}\chi=\varepsilon_{\mu\nu}\hat{J}^{\nu
}(t)\chi-t\hat{J}_{\mu}(t)\chi +t\partial_{\mu}\chi\,.  \label{eq}
\end{equation}
Using the form of $\hat{J}_{\mu} (t)$ given in~(\ref{flatcurrent}), we
can write~(\ref{eq}) as
\begin{eqnarray}
&& \varepsilon_{\mu\nu}\left( \partial^{\nu}-H^{\nu}-P^{\nu}-Q^{\nu}\right)
\chi=
\\&&
\quad=t\left(
\partial_{\mu}-H_{\mu}\!+\!P_{\mu}-\frac{1}{\sqrt{1\!-\!t^{2}}}Q_{\mu }\!-\!
\frac{t}{\sqrt{1-t^{2}}}\left( Q^{^{\prime}}\right) _{\mu} \!+\!\frac{1}
     {\sqrt{ 1\!-\!t^{2}}}\varepsilon_{\mu\nu}\left( Q^{^{\prime}}\right)
     ^{\nu }+S(t)\varepsilon_{\mu\nu}Q^{\nu}\right) \chi, \label{main}
\nonumber
\end{eqnarray}
where $S(t)$ can be determined from
\begin{equation}
\frac{1}{\sqrt{1-t^{2}}}=1+tS(t)\,.  \label{dec1}
\end{equation}

Let us note that  
\begin{equation}
H^{\mu}+P^{\mu}+Q^{\mu}=-\hat{g}^{-1} (0)\partial^{\mu}\hat{g} (0) =
g^{-1} \partial^{\mu} g = J^{\mu}(0)\,, \label{V0}
\end{equation}
so that we can write~(\ref{main}) in the form
\begin{eqnarray}
&& \varepsilon_{\mu\nu}g^{-1}\partial^{\nu}\left(
  g\chi\right)=
\\
&&
=t\left(\partial_{\mu}\!-\!H_{\mu}+P_{\mu}\!-\!\frac{1}{\sqrt{1-t^{2}}}Q_{\mu}\!-\!
\frac {t}{\sqrt{1\!-\!t^{2}}}\left( Q^{^{\prime}}\right) _{\mu}
+\frac{1}{\sqrt {1-t^{2}}} \varepsilon_{\mu\nu}\left(
Q^{^{\prime}}\right) ^{\nu
}+S(t)\varepsilon_{\mu\nu}Q^{\nu}\right)\chi\,.  \label{m2}
\nonumber
\end{eqnarray}
We can now expand $\chi$ in a power series in the spectral parameter
as
\begin{equation}
\chi=\overset{\infty}{\underset{n=0}{\sum}}\ t^{n}\chi^{(n-1)}\,.  \label{de1}
\end{equation}
Substituting this into~(\ref{m2}), we obtain
\begin{eqnarray}
\overset{\infty}{\underset{n=0}{\sum}}\
  t^{n}\varepsilon_{\mu\nu}g^{-1}\partial^{\nu}\left(g\chi^{(n-1)}\right)&=&
 \overset{\infty}{\underset{n=0}{\sum}}\
  t^{(n+1)}\biggl\{\partial_{\mu}- H_{\mu
}+P_{\mu}-\frac{1}{\sqrt{1-t^{2}}}Q_{\mu}-\frac{t}{\sqrt{1-t^{2}}}\left(
Q^{^{\prime}}\right) _{\mu}+
\nonumber\\&&
         \hphantom{\overset{\infty}{\underset{n=0}{\sum}}\
  t^{(n+1)}\biggl\{}
+\frac{1}{\sqrt{1-t^{2}}}\varepsilon_{\mu\nu}\left(
  Q^{^{\prime}}\right) 
^{\nu}+S(t)\varepsilon_{\mu\nu}Q^{\nu}\biggr\}\chi^{(n-1)}\,. \label{m3}
\end{eqnarray}
Furthermore, expanding $\frac{1}{\sqrt{1-t^{2}}}$ as a power series
\begin{equation}
\frac{1}{\sqrt{1-t^{2}}}=\allowbreak1+\frac{1}{2}t^{2}+\frac{3}{8}
t^{4}+ \cdots = \sum_{n}\ r_{n}t^{n}\,, \label{de2}
\end{equation}
we can iteratively determine all the $\chi^{(n)}$'s from~(\ref{m3}).

We note that the lowest order term in powers of $t$ (namely, $t^{0}$)
in~(\ref{m3}) gives
\begin{eqnarray}
\varepsilon_{\mu\nu}g^{-1}\partial^{\nu}\left( g\chi^{(-1)}\right) &=&0
\label{1}\\
{\rm or,} \qquad
g^{-1}\partial^{\nu}g\chi^{(-1)}+\partial^{\nu}\chi^{(-1)}&=&0\,.
\label{2}
\end{eqnarray}
This implies that $g\chi^{(-1)}$ is a constant and it is convenient to
choose
\begin{equation}
g\chi^{(-1)}=\frac{1}{2}\,, \label{set}
\end{equation}
for later purposes. In the linear order in $t$, Eq.~(\ref{m3}) leads
to (after using~(\ref{2}))
\begin{eqnarray}
\varepsilon_{\mu\nu}g^{-1}\partial^{\nu}\left( g\chi^{(0)}\right) 
&=&2\left( P_{\mu}+\frac{1}{2}\varepsilon_{\mu\nu}(Q^{\prime
})^{\nu}\right) \chi^{(-1)}  
\label{t1} \\
& =&\left( P_{\mu}+\frac{1}{2}\varepsilon_{\mu\nu}(Q^{\prime}
)^{\nu}\right) g^{-1}\,,  \label{t11}
\end{eqnarray}
and so on.

We can now define the conserved currents associated with the theory as
\begin{equation}
j_{\mu}^{(n)}=\varepsilon_{\mu\nu}\partial^{\nu}\left( g\chi
^{(n)}\right).  \label{cur2}
\end{equation}
This leads to the identification
\begin{eqnarray}
j_{\mu}^{(-1)} & = &
\varepsilon_{\mu\nu}\partial^{\nu}\left(g\chi^{(-1)}\right) =
0,\nonumber\\ 
j_{\mu}^{(0)} & = & \varepsilon_{\mu\nu}\partial^{\nu}\left(g
\chi^{(0)}\right) =g\left(
P_{\mu}+\frac{1}{2}\varepsilon_{\mu\nu}(Q^{\prime})^{\nu}\right)
g^{-1}\,.\label{j0}
\end{eqnarray}
Upon using~(\ref{convrule}), the second current takes the form
\begin{equation}
j_{\mu}^{(0)}=
p_{\mu}+\frac{1}{2}\varepsilon_{\mu\nu}(q^{\prime})^{\nu}\,.\label{final}
\end{equation}
This is precisely the Noether current~(\ref{noether}) determined
earlier which is conserved.  Similarly, in the second order in $t$, we
obtain
\begin{eqnarray}
j_{\mu}^{(1)} & =& \varepsilon_{\mu\nu}\partial^{\nu}\left(
g\chi^{(1)}\right)  
\nonumber\\
& =&\varepsilon_{\mu\nu}\left(j^{(0)}\right) ^{\nu}+2j_{\mu}^{(0)}\left[
g\chi^{(0)}\right] +\frac{1}{4}\varepsilon_{\mu\nu}q^{\nu}-\frac{1}{2}
q_{\mu}^{^{\prime}}\,,  \label{j1}
\end{eqnarray}
which can also be written in an explicitly non-local form
using~(\ref{j0}) as
\begin{equation}
j_{\mu}^{(1)} = \varepsilon_{\mu\nu}\left(j^{(0)}\right)
^{\nu}+2j_{\mu}^{(0)}\left(\partial^{-1}j_{0}^{(0)}\right)
+\frac{1}{4}\varepsilon_{\mu\nu}q^{\nu}-\frac{1}{2}
q_{\mu}^{^{\prime}}\,.
\end{equation}
It is easy to check (using the form~(\ref{j1})) that this current is
indeed conserved, $\partial^{\mu}j_{\mu}^{(1)}=0$. Using this
iterative procedure we can construct all the conserved currents in the
hierarchy which are left invariant. The corresponding algebra of the
non-local charges is expected to satisfy a yangian
algebra~\cite{mackay, KEK}.

\section{The hamiltonian analysis}\label{section6}

The calculation of the algebra of charges can be carried out once the
basic Poisson brackets of the theory have been determined. Basically,
we are interested in the Poisson algebra of the transition
matrices. Let us recall that the transition matrix
$T(\sigma_{1},\sigma_{2},t)$ is defined using the
current~(\ref{flatcurrent}) which satisfies the zero curvature
condition as
\begin{equation}
T(\sigma_{1},\sigma_{2},t)=g^{-1}(\sigma_{1},t)g(\sigma_{2},t) = {\rm
  P}\left(e^{\int_{\sigma_{2}}^{\sigma_{1}} \mathrm{d}\sigma\
  \hat{J}_{1} (\sigma,t)}\right).  \label{trans}
\end{equation}
Here, we have put back the explicit dependence on the worldsheet
coordinates whose spatial component is denoted by $\sigma$. (The
transition matrix is simply an open Wilson line along a spatial
path. The spatial coordinate $\sigma$ is periodic for the
$AdS_{5}\times S^{5}$ string and this causes some technicalities in
defining the charges. We avoid such questions by working directly with
the transition matrix.)  It follows now~\cite{FT, das} that
\begin{eqnarray}
\{\overset{1}{T}(\sigma_{1},\sigma_{2},t_{1}),\overset{2}{T}(\sigma_{1}^
{\prime}, \sigma_{2}^{\prime},t_{2})\} 
&=&\int_{\sigma_{2}}^{\sigma_{1}}\mathrm{d}\sigma 
\int_{\sigma_{2}^{\prime}}^{\sigma_{1}^{\prime}}
\mathrm{d}\sigma^{\prime}\left(\overset{1}{T}(\sigma_{1},\sigma,t_{1}),
\overset{2}{T}
(\sigma_{1}^{\prime},\sigma^{\prime},t_{2})\right)\times
\nonumber\\&&
\times\,\left\{\overset{1}{\hat{J}_{1}}(\sigma,t_{1}),\overset{2}{\hat{J}_{1}}(\sigma^
{\prime},t_{2})\right\}\left(\overset{1}{T}(\sigma,\sigma_{2},t_{1}),\overset{2}{T}
(\sigma^{\prime},\sigma_{2}^{\prime},t_{2})\right).  
\label{fad}
\end{eqnarray}
We note here that~(\ref{fad}) represents a matrix Poisson bracket
written in an index free tensor notation~\cite{FT} (which we will
follow in our analysis) defined as
\begin{equation}
\overset{1}{A}=A\otimes I\,,\qquad \overset{2}{B}=I\otimes B,  \label{a1b2}
\end{equation}
where 
\begin{equation}
(A\otimes B)_{ij,km}=A_{ik}B_{jm}\,.  \label{tensor}
\end{equation}
It follows from~(\ref{a1b2}) and~(\ref{tensor}) that
\begin{equation}
(A\otimes B)(C\otimes D)=(-)^{\epsilon_{B}\epsilon_{C}}\left(
AC\otimes BD\right).  \label{rule1}
\end{equation}
Furthermore, from~(\ref{fad}), we see that in order to compute the
Poisson bracket between the transition matrix, it is necessary to
evaluate $\{\overset{1}{\hat{J}_{1}}(\sigma,t_{1}),\overset{2
}{\hat{J}_{1}}(\sigma^{\prime},t_{2})\}$ which we can do only after
carrying out a hamiltonian analysis of the model.

The hamiltonian analysis~\cite{HENTEI} can be carried out starting
from the action~(\ref{newaction}) in a way similar to~\cite{KS}. We
treat the space component of the current $J_{\mu}$ as the dynamical
variable. The zero curvature condition~(\ref{zc}) then allows us to
determine the time component of the current as
\begin{equation}
J_{0}=D^{-1}\left(\partial_{0}J_{1}\right), \label{zerocomp}
\end{equation}
where (we are identifying
$\partial_{0}=\frac{\partial}{\partial\tau},\partial_{1}=
\frac{\partial}{\partial\sigma}$ corresponding to the two worldsheet
coordinates)
\begin{equation}
D = \partial_{1}-[J_{1},\cdot]\,.  \label{covariantder}
\end{equation}
The canonical momentum can now be obtained from an arbitrary variation
of the action~(\ref{newaction}) satisfying~(\ref{zerocomp}) and leads
to (we use the left derivatives for fermionic degrees of freedom)
\begin{equation}
\Pi_{J}\equiv\Pi_{H}\oplus\Pi_{Q^{1}}\oplus\Pi_{P}\oplus\Pi_{Q^{2}}
=-D^{-1}\left( P_{0}+\frac{1}{2}Q_{1}^{1}-\frac{1}{2}Q_{1}^{2}\right).
\label{momenta}
\end{equation}
In components (see section~\ref{section3}), the basic canonical
Poisson bracket structures are given by (at equal time)
\begin{eqnarray}
\{P_{1}^{a}(\sigma),\left( \Pi_{P}\right) _{b}(\sigma^{\prime})\} 
&=&\delta_{b}^{a} \delta (\sigma - \sigma^{\prime})\,,
\label{Pp} \\
\{H_{1}^{m}(\sigma),\left( \Pi_{H}\right) _{n}(\sigma^{\prime})\} & =&
\delta_{n}^{m} \delta (\sigma - \sigma^{\prime})\,,
\label{Hh} \\
\{\left( Q_{1}^{1}\right)^{\alpha_{1}}(\sigma),\left(\Pi_{Q^{2}}\right)
_{\beta_{1}}(\sigma^{\prime})\} 
&=&-\delta_{\beta_{1}}^{\alpha_{1}}\delta (\sigma - \sigma^{\prime})\,,
\label{Q1q2} \\
\{\left( Q_{1}^{2}\right) ^{\alpha_{2}}(\sigma),\left( \Pi_{Q^{1}}\right)
_{\beta_{2}}(\sigma^{\prime})\} 
&=&-\delta_{\beta_{2}}^{\alpha_{2}}\delta (\sigma - \sigma^{\prime})\,.
\label{Q2q1}
\end{eqnarray}
The extra minus sign in~(\ref{Q1q2}) and~(\ref{Q2q1}) arises as a
result of the definition of the generalized Poisson brackets involving
fermionic systems~\cite{HENTEI}
\begin{equation}
\{F,G\}=\left[ \frac{\partial F}{\partial q^{i}}\frac{\partial
    G}{\partial \pi_{i}}-\frac{\partial F}{\partial
    \pi_{i}}\frac{\partial G}{\partial q^{i}} \right]
+(-)^{\epsilon_{F}}\left[\frac{\partial^{L}F}{\partial\theta^{\alpha}}\frac
  {\partial^{L}G}{\partial\pi_{\alpha}}+\frac{\partial^{L}F}
  {\partial\pi_{\alpha}}\frac
  {\partial^{L}G}{\partial\theta^{\alpha}}\right], \label{genpoisson}
\end{equation}
where the superscript $L$ represents left derivation while
$\epsilon_{F}$ denotes Grassmann parity of $F$.

Decomposing relation~(\ref{momenta}) into the appropriate subspaces,
we obtain
\begin{eqnarray}
P_{0} & =&-\partial_{1}\Pi_{P}+[H_{1},\Pi_{P}]+[P_{1},\Pi_{H}]+[Q_{1}^{1}
,\Pi_{Q^{1}}]+[Q_{1}^{2},\Pi_{Q^{2}}]\,,  
\label{P0constraint} \\
\varphi_{1} & =& -\partial_{1}\Pi_{H}+[H_{1},\Pi_{H}]+[P_{1},\Pi_{P}]+[Q_{1}^{1}
,\Pi_{Q^{2}}]+[Q_{1}^{2},\Pi_{Q^{1}}] \approx 0\,,  
\label{Hconstraint}\\
\varphi_{2} & =&-\frac{1}{2}Q_{1}^{1}-\partial_{1}\Pi_{Q^{1}}+[H_{1},\Pi_{Q^{1}
}]+[P_{1},\Pi_{Q^{2}}]+[Q_{1}^{1},\Pi_{H}]+[Q_{1}^{2},\Pi_{P}] \approx
0\,,
\label{Q1constraint} \\
\varphi_{3} & =&\frac{1}{2}Q_{1}^{2}-\partial_{1}\Pi_{Q^{2}}+[H_{1},\Pi_{Q^{2}}
]+[P_{1},\Pi_{Q^{1}}]+[Q_{1}^{1},\Pi_{P}]+[Q_{1}^{2},\Pi_{H}] \approx
0\,.   \label{Q2constraint}
\end{eqnarray}
Since $P_{0}$ contains time derivatives (see~(\ref{zerocomp})), the
first relation can be used to express velocities in terms of canonical
momenta. The last three relations, on the other hand, define primary
constraints of the theory. In particular, the first
constraint~(\ref{Hconstraint}) is the generator of gauge
transformation~(\ref{gaugetrans}). On the other hand, as we will see
the last two fermionic constraints give rise to the
$\kappa$-symmetry. The three primary
constraints~(\ref{Hconstraint}),~(\ref{Q1constraint})
and~(\ref{Q2constraint}) should be supplemented with the standard
Virasoro constraints, which in our notation take the forms
\begin{eqnarray}
\varphi_{4} &=& \frac{1}{2}\ \mbox{str}(P_{0}^{2}+P_{1}^{2}) \approx 0\,,
\label{1stVirasoro}  \\
\varphi_{5} &=& \mbox{str}(P_{0}P_{1})  \approx 0\,,  \label{2ndVirasoro}
\end{eqnarray}
and can also be written as
\begin{equation}
\varphi_{\pm} = {\rm str}\left(P_{0}\pm P_{1}\right)^{2} \approx 0\,.
\end{equation}

The Poisson brackets~(\ref{Pp}),~(\ref{Hh}),~(\ref{Q1q2}) and~(\ref{Q2q1}) can be written in the index free tensor notation as
\begin{eqnarray}
\left\{\overset{1}{P}_{1}(\sigma),\overset{2}{\Pi}_{P}(\sigma^{\prime})\right\} &
=& \Omega_{P} \delta (\sigma - \sigma^{\prime})\,, 
\nonumber\\
\left\{\overset{1}{H}_{1}(\sigma),\overset{2}{\Pi}_{H}(\sigma^{\prime})\right\} &
=& \Omega_{H} \delta (\sigma - \sigma^{\prime}),
\nonumber\\
\left\{\overset{1}{Q_{1}^{1}}(\sigma),\overset{2}{\Pi}_{Q^{2}}(\sigma^{\prime})\right\}
& =& \Omega_{Q^{12}}\delta (\sigma - \sigma^{\prime})\,,  
\nonumber \\
\left\{\overset{1}{Q_{1}^{2}}(\sigma),\overset{2}{\Pi}_{Q^{1}}(\sigma^{\prime})\right\}
& =& \Omega_{Q^{21}}\delta (\sigma - \sigma^{\prime})\,,
\label{tensorpoisson} 
\end{eqnarray}
where we have introduced the Casimir operators 
\begin{eqnarray}
\Omega_{P}& = & t_{p}\otimes
t_{p^{\prime}}G^{pp^{\prime}}\,, \qquad\Omega_{H}=t_{h}\otimes t_{h^{\prime}
}G^{hh^{\prime}}\,,\nonumber\\
\Omega_{Q^{12}} & = & t_{\alpha_{1}}\otimes t_{\alpha_{2}
}G^{\alpha_{1}\alpha_{2}}\,,\qquad \Omega_{Q^{21}}=t_{\alpha_{2}}\otimes
t_{\alpha_{1}}G^{\alpha_{2}\alpha_{1}}\,.
\end{eqnarray} 
It can now be derived in a straightforward manner (using the
relations~(\ref{a1}) presented in the \textref{section8}{appendix})
that
\begin{eqnarray}
\left\{\overset{1}{P}_{0}(\sigma),\overset{2}{H}_{1}(\sigma^{\prime})\right\} & =&
-\left[\Omega_{P},\overset{2}{P}_{1}(\sigma)\right]\delta (\sigma - \sigma^{\prime}),
\nonumber \\[4pt]
\left\{\overset{1}{P}_{0}(\sigma),\overset{2}{Q_{1}^{1}}(\sigma^{\prime})\right\}
& =& -\left[\Omega_{P} ,\overset{2}{Q_{1}^{2}}(\sigma)\right]\delta (\sigma -
\sigma^{\prime})\,,   
\nonumber \\[4pt]
\left\{\overset{1}{P}_{0}(\sigma),\overset{2}{Q_{1}^{2}}(\sigma^{\prime})\right\}
& =& -\left[\Omega_{P} ,\overset{2}{Q_{1}^{1}}(\sigma)\right]\delta (\sigma -
\sigma^{\prime})\,,   \nonumber \\[4pt]
\left\{\overset{1}{P}_{0}(\sigma),\overset{2}{P}_{1}(\sigma^{\prime})\right\} & =&
-\left[\Omega_{P},\overset{2}{H}_{1}(\sigma)\right]\delta (\sigma -
\sigma^{\prime})+\Omega_{P}\partial_{\sigma}\delta (\sigma -
\sigma^{\prime})\,,   \nonumber \\[4pt]
\left\{\overset{1}{P}_{0}(\sigma),\overset{2}{P}_{0}(\sigma^{\prime})\right\} & =&
-\left[\Omega_{P},\overset{2}{\varphi}_{1}(\sigma)\right]\delta (\sigma -
\sigma^{\prime})\,.  \label{rel1}
\end{eqnarray}

\looseness=-1 We note that there is a non-ultra local term in the Poisson
brackets~(\ref{rel1}) (namely, the term with the derivative acting on
the delta function which in the context of integrable systems is
called a non-ultralocal term, while it is known as a Schwinger term in
field theory. We will use these two terms interchangeably.). It is
well known that the presence of such a term leads to problems in the
calculation of the Poisson brackets between the transition
matrices. In the bosonic case considered in~\cite{KS}, this
problematic term is naturally regularized in the calculation of the
algebra of transition matrices in the presence of the dilaton
field. When there is no coupling to gravity (and hence a constant
spectral parameter), there are also several methods of regularizing
the calculations and we will discuss some of them later. We note that
the presence of $\kappa$ symmetry, in the present case, also
complicates the calculations. The next step in our analysis is,
therefore, to determine all the secondary constraints of the theory
and group the constraints into first class and second class
constraints. The total hamiltonian density of the theory is easily
determined to be
\begin{equation}
\mathcal{H}_{T}= {\rm
  str}\left(\frac{1}{2}\left(P_{0}^{2}+P_{1}^{2}\right) + \lambda
_{1}\varphi_{1}+\lambda_{2}\varphi_{2}+\lambda_{3}\varphi_{3}\right),
\label{hamiltonain}
\end{equation}
where $\lambda_{1},$ $\lambda_{2}$ and $\lambda_{3}$ denote the
Lagrange multipliers corresponding to the three primary
constraints. $\varphi_{1}$ is easily checked to be stationary while
requiring the constraints $\varphi_{2},\varphi_{3}$ to be stationary
determines two of the Lagrange multipliers to correspond to
\begin{equation}
\lambda_{2}=-Q_{1}^{2},\quad\lambda_{3}= Q_{1}^{1}\,.  \label{lam23}
\end{equation}
There are no additional secondary constraints and the Lagrange
multiplier $\lambda_{1}$ is undetermined corresponding to the fact
that $\varphi_{1}$ is the generator of a gauge
symmetry. $\varphi_{4},\varphi_{5}$ or equivalently $\varphi_{\pm}$
can also be checked to be conserved under the hamiltonian flow. It is
important to emphasize here that unlike in bosonic models where the
standard Virasoro constraints correspond to first class constraints
representing generators of reparameterization transformations, in the
present theory (in this formulation) with fermions, this is not true.
On the other hand, one can define a linear combination of the
constraints as
\begin{eqnarray}
\bar{\varphi}_{4} & = & \varphi_{4} + {\rm str}\
\left(\lambda_{2}\varphi_{2} + \lambda_{3}\varphi_{3}\right) \approx
0\,,
\nonumber\\ 
\bar{\varphi}_{5} & = & \varphi_{5} - {\rm str}\
\left(\lambda_{2}\varphi_{2} - \lambda_{3}\varphi_{3}\right) \approx
0\,,
\end{eqnarray}
which can be easily checked to correspond to first class constraints
and generate the reparameterization transformation (of course,
$\bar{\varphi}_{4}$ corresponds to the hamiltonian).

It is straightforward to calculate the Poisson brackets between the
constraints
\begin{eqnarray}
\left\{\overset{1}{\varphi}_{1}(\sigma),\overset{2}{\varphi}_{1}(\sigma^{\prime})\right\}
& =& -\left[\Omega _{H},\overset{2}{\varphi}_{1}(\sigma)\right]\delta (\sigma -
\sigma^{\prime}) \approx 0\,,  
\nonumber \\
\left\{\overset{1}{\varphi}_{1}(\sigma),\overset{2}{\varphi}_{2}(\sigma^{\prime})\right\}
& =& -\left[\Omega _{H},\overset{2}{\varphi}_{2}(\sigma)\right]\delta (\sigma -
\sigma^{\prime}) \approx 0\,,  
\nonumber \\
\left\{\overset{1}{\varphi}_{1}(\sigma),\overset{2}{\varphi}_{3}(\sigma^{\prime})\right\}
& =& -\left[\Omega _{H},\overset{2}{\varphi}_{3}(\sigma)\right]\delta (\sigma -
\sigma^{\prime}) \approx 0\,.  \label{fi1fi23}
\end{eqnarray}
This shows explicitly that $\varphi_{1}$ is indeed a first class
constraint as we expect. The algebra between $\varphi_{2}$ and
$\varphi_{3}$, on the other hand, is more complicated
\begin{eqnarray}
\left\{\overset{1}{\varphi}_{2}(\sigma),\overset{2}{\varphi}_{2}(\sigma^{\prime})\right\}
& =& -\left[\Omega_{Q^{12}},\overset{\left( 2\right) }{\left( P_{0}-P_{1}\right)}
(\sigma)\right]\delta (\sigma - \sigma^{\prime})\,,  \label{fi2fi2} \\
\left\{\overset{1}{\varphi}_{2}(\sigma),\overset{2}{\varphi}_{3}(\sigma^{\prime})\right\}
& =& -\left[\Omega_{Q^{12}},\overset{2}{\varphi}_{1}(\sigma)\right]\delta (\sigma
- \sigma^{\prime}) \approx 0\,,   \label{fi2fi3} \\
\left\{\overset{1}{\varphi}_{3}(\sigma),\overset{2}{\varphi}_{3}(\sigma^{\prime})\right\}
& =& -\left[\Omega_{Q^{21}},\overset{\left( 2\right) }{\left( P_{0}+P_{1}\right) }
(\sigma)\right]\delta (\sigma - \sigma^{\prime})\,.  \label{fi3fi3}
\end{eqnarray}
We see from~(\ref{fi2fi2}) and~(\ref{fi3fi3}) that $\varphi_{2}$ and $
\varphi_{3}$ define a non trivial algebra. These constraints are,
however, reducible because of the constraints~(\ref{1stVirasoro})
and~(\ref{2ndVirasoro}). One, therefore, has to further decompose
$\varphi_{2}$ and $\varphi _{3}$ into first and second class
constraints using some relevant projection~\cite{KEK}. The resulting
first class constraints will then generate the $\kappa$-symmetry. The
second class constraints can be used to define the Dirac brackets. The
decomposition of $\varphi_{2},\varphi_{3}$ into first class and second
class components, however, is nontrivial and remains an open
question. In this paper, we attempt to calculate the ordinary Poisson
bracket between the currents which can be thought of as a first step
in the complete evaluation of the algebra of the transition matrices.

We can calculate the desired
$\{\overset{1}{\hat{J}_{1}}(\sigma,t_{1}),\overset{2}{\hat{J}_{1}}
(\sigma^{\prime},t_{2})\}$ bracket using~(\ref{flatcurrent}), the
relations (\ref{tensorpoisson}) as well as~(\ref{rel1}). Without
giving the tedious technical details, we note that the form of this
bracket takes the closed form
\begin{eqnarray}
&& \left\{\overset{1}{\hat{J}_{1}}(\sigma,t_{1}),\overset{2}{\hat{J}_{1}}
(\sigma^{\prime}, t_{2})\right\}=
\nonumber\\&&
\quad =\left(
\alpha\left[\Omega_{P},\overset{\left(1\right) }{\hat{J}_{1}}
(\sigma,t_{1})\right]+\beta\left[\Omega_{P},\overset{\left(2\right) }{\hat{J}
_{1}}(\sigma,t_{2})\right]+\gamma\left[\Omega_{H},\overset{\left(1\right)}
{\hat{J}_{1}}(\sigma,t_{1})+\overset{\left(2\right)
 }{\hat{J}_{1}}(\sigma,t_{2})\right]\right)
\delta (\sigma - \sigma^{\prime}) + 
\nonumber\\&&
         \hphantom{\quad =}
+\xi_{2}\left\{\overset{1}{\hat{J}_{1}}(\sigma,t_{1}),\overset{2}
{\varphi}_{2}(\sigma^{\prime})\right\}+\chi_{2}\left\{\overset{1}{\hat{J}_{1}}
(\sigma,t_{1}),\overset{2}{\varphi}_{3}(\sigma^{\prime})\right\}+
\nonumber\\&&
         \hphantom{\quad =}
 +\xi_{1}\left\{\overset{1}{\varphi}_{2}(\sigma^{\prime}),\overset{2}{\hat{J}_{1}}
 (\sigma,t_{1})\right\}+\chi_{1}\left\{\overset{1}{\varphi}_{3}(\sigma^{\prime}),
 \overset{2} {\hat{J}_{1}} (\sigma,t_{1})\right\}
 +\Lambda\partial_{\sigma}\delta (\sigma - \sigma^{\prime})\,,
\label{central}
\end{eqnarray}
where $\alpha,\beta,\gamma,\xi_{1,2}$ and $\chi_{1,2}$ are 
functions of the spectral parameters $t_{1},t_{2}$ defined as
\begin{eqnarray}
\alpha & =&\frac{(B_{2})^{2}}{A_{1}B_{2}-A_{2}B_{1}}\,,\qquad \beta
=\alpha\left( \frac{B_{1}}{B_{2}}\right) ^{2}\,,\qquad \gamma=\alpha \frac{
B_{1}}{B_{2}}\,,
\nonumber \\
\xi_{2} & =&\gamma C_{2}\,,\qquad \chi_{2}=\gamma D_{2}\,,\qquad \xi
_{1}=-\gamma C_{1}\,,\qquad \chi_{1}=-\gamma D_{1}\,,  
\label{coef1} \\
\Lambda & =& \left( A_{1}B_{2}+A_{2}B_{1}\right) \Omega_{P}+\left( \xi
_{2}D_{1}+\chi_{1}C_{2}\right) \Omega_{Q^{21}}  +
\nonumber\\&&
+\,\left( \xi_{1}D_{2}+\chi_{2}C_{1}\right) \Omega_{Q^{12}}\,,
\end{eqnarray}
with
\begin{equation}
A_{i}=\frac{1+t_{i}^{2}}{1-t_{i}^{2}},\text{ }B_{i}=\frac{2t_{i}}{
  1-t_{i}^{2} },\text{ }C_{i}=\sqrt{\frac{1+t_{i}}{1-t_{i}}},\text{
}D_{i}= \sqrt {\frac{1-t_{i}}{1+t_{i}}}\,.  \label{abcd}
\end{equation}

In the bosonic limit, i.e.\ when one sets all the fermions to zero,
this reduces to the result of~\cite{KS}. In the presence of the
fermions one has additional terms depending on $\xi_{i},\chi_{i}$ as
well as non-ultralocal terms involving $\Lambda$. The terms depending
on $\xi_{i},\chi_{i}$ are there primarily because we have not yet
separated the constraints into ($\kappa$ symmetry generating) first
class constraints and second class constraints. Once this is done and
the second class constraints are used to define Dirac brackets, then,
in the Dirac bracket of the currents, such terms will be absent and
the algebra will have a closed structure. The $\kappa$ symmetry can,
in fact, be fixed in the action as has been suggested in~\cite{kr2,
  KalTs, pes1, pes2}. It is also interesting to understand the meaning
of the algebra of the transition matrices on the SYM side. This is a
topic presently under study.

\looseness=1 The presence of the $\Lambda$ dependent non-ultralocal
terms, on the other hand, leads to a different issue. Analogous to the
bosonic case considered in~\cite{KS}, one has to deal with the fact
that the non-ultralocal term in~(\ref{rel1}) will lead to an ambiguity
in the calculation of the bracket between the transition
matrix~(\ref{fad}). There have been several methods proposed to
regularize this ambiguity for the PCM (principal chiral model) and
other models. They are based either on regularizing~\cite{Maillet} the
Poisson bracket between the currents~(\ref{central}) by defining
symmetrized ``weak" Poisson brackets, or by regularizing the
transition matrices by introducing a "retarded" monodromy
matrix~\cite{NIDUN}. Another method due to Faddeev and
Reshetikhin~\cite{FARESH} views the non-ultralocal terms as a
consequence of false vacuum in the classical limit and correspondingly
modify the vacuum structure of the theory.

\section{Summary and discussions}\label{section7}

In this paper, we have constructed nonlocal conserved currents for the
superstring in $AdS_5 \times S_5$ background, investigated the
hamiltonian structure and presented the algebra of the transition
matrices.  It is noted that the evolution of the superstring in $AdS_5
\times S_5$ is most conveniently described in Green-Schwarz formalism
and the action is expressed as a nonlinear $\sigma$ model on the coset
$\frac{\PSU(2,2|4)}{\SO(4,1) \times \SO(5)}$. It is pointed out that
superalgebra admits a $Z_4$ automorphism which determines the maximal
subgroup as $\SO(4,1) \times \SO(5)$. The model naturally contains a
current satisfying the zero curvature condition.

We introduce a constant spectral parameter which enables us to define
a family of new currents such that the vanishing of curvature of these
new currents leads to the equations of motion as well as the flatness
of the original current.  These constructed currents, in our
formulation, are invariant under definite transformation rules which
are generalizations of the $Z_4$ automorphism of the
superalgebra. This is completely parallel to the bosonic sigma models
where this generalized automorphism is used in the construction of the
monodromy matrix and the nonlocal currents.

\looseness=1 It is more appropriate to adopt the hamiltonian formalism to compute
the Poisson bracket algebra of the transition matrices (which can
yield the algebra of the charges). The hamiltonian analysis is carried
out in a frame work where the spatial component of the current is
identified as dynamical variables and its time component gets
determined from the zero curvature condition.  Subsequently, the
canonical momenta are identified. Next, the constrained hamiltonian
analysis is carried out to identify the constraints and the algebra of
the constraints is presented. When we compute the Poisson bracket
between the new currents (defined with the spectral parameter) there
are additional terms which include Schwinger term (derivative of the
$\delta$-function). As a cross check, if we set all fermionic
coordinates to zero, we are able to recover the algebra of such
currents derived earlier for purely bosonic $\sigma$-models. However,
the presence of the $\kappa$ symmetry makes it difficult to separate
the constraints into first and second class ones and construct the
Dirac brackets. It is necessary, therefore, to fix the $\kappa$
symmetry, this is presently under study.

The presence of the non-ultralocal term in the algebra of the current
leads to ambiguities in the computation of brackets between the
transition matrices. There already exist proposals to regularize such
terms as we have discused earlier. The essential difference between
the bosonic sigma model coupled to gravity and the present theory is
that here we have a constant spectral parameter. We note, however,
that in the present theory we do have a $\mathbb{Z}_{4}^{\infty}$
invariance~(\ref{symmet3}) that leads to a symmetry under
$t\longrightarrow\frac{1}{t}$ much like in the bosonic case. This
naturally suggests that one way to regularize the non-ultralocal terms
is to assume that the spectral parameter is a local function
satisfying~(\ref{spectcondition}) (which would correspond to having a
dilaton field in the theory). This would naturally
regularize~\cite{KS} the ambiguity in~(\ref{fad}) and only at the end
of the calculation one should take the limit of a constant spectral
parameter. This is an interesting possibility that needs further work
and is under investigation.

\acknowledgments

One of us (M.S.) would like to thank Kentaroh Yoshida for useful
discussions. This work is supported in part by US DOE Grant No.~DE-FG
02-91ER40685.

\appendix

\section{Useful identities}\label{section8}

We present here some relations that have proved useful in the text. 
\begin{equation}
\begin{array}[b]{rclcrcl}
\left[ \Omega _{H},\overset{1,2}{P}\right]&=&-\left[\Omega _{P},\overset{2,1}{P}\right],
&\qquad& \left[\Omega _{P},\overset{1}{H}\right]&=&-\left[\Omega _{P},\overset{2}{H}\right],  \\[7pt]
\left[ \Omega _{Q^{12}},\overset{2}{Q^{2}}\right]&=&-\left[\Omega 
_{P},\overset{1}{
Q^{2}}\right],&\qquad& \left[\Omega _{Q^{21}},\overset{2}{Q^{1}}\right]&=&-\left[\Omega 
_{P},\overset{1}{Q^{1}}\right],  \\[7pt]
\left[ \Omega _{Q^{21}},\overset{1}{Q^{2}}\right]&=&\left[\Omega _{P},\overset{2}{Q^{2}
}\right],&\qquad& \left[\Omega _{Q^{12}},\overset{1}{Q^{1}}\right]&=&\left[\Omega 
_{P},\overset{2}{Q^{1}}\right], \\[7pt]
\left[ \Omega _{Q^{12}},\overset{2}{Q^{1}}\right]&=&-\left[\Omega _{H},\overset{1}{
Q^{1}}\right],&\qquad& \left[\Omega _{Q^{21}},\overset{2}{Q^{2}}\right]&=&-\left[\Omega 
_{H},\overset{1}{Q^{2}}\right],   \\[7pt]
\left[ \Omega _{Q^{12}},\overset{1}{Q^{2}}\right]&=&\left[\Omega _{H},\overset{2}{Q^{2}
}\right],&\qquad& \left[\Omega _{Q^{21}},\overset{1}{Q^{1}}\right]&=&\left[\Omega 
_{H},\overset{2}{Q^{1}}\right], \\[7pt]
\left[ \Omega _{Q^{12}},\overset{1}{H}\right]&=&-\left[\Omega _{Q^{12}},\overset{2}{H}
\right],&\qquad& \left[\Omega _{Q^{21}},\overset{1}{H}\right]&=&-\left[\Omega 
_{Q^{21}},\overset{2}{H}\right], \\[7pt]
\left[ \Omega _{Q^{12}},\overset{1}{P}\right]&=&-\left[\Omega _{Q^{21}},\overset{2}{P}
\right],&\qquad& \left[\Omega _{Q^{12}},\overset{2}{P}\right]&=&-\left[\Omega 
_{Q^{21}},\overset{1}{P}\right], \\[7pt]
\left[ \Omega _{H},\overset{1}{H}\right]&=&-\left[\Omega _{H},\overset{2}{H}\right].  &\qquad& {}&{}
\label{a1}
\end{array}
\end{equation}


\begin{thebibliography}{99}

\bibitem{mald}
J.M.~Maldacena, \emph{The large-$N$ limit of superconformal field
  theories and supergravity}, \atmp{2}{1998}{231} [\hepth{9711200}].

\bibitem{kleb}
S.S.~Gubser, I.R.~Klebanov and A.M.~Polyakov, \emph{Gauge theory
  correlators from non-critical string theory}, \plb{428}{1998}{105}
[\hepth{9802109}].

\bibitem{witt}
E.~Witten, \emph{Anti-de~Sitter space and holography},
\atmp{2}{1998}{253} [\hepth{9802150}].

\bibitem{rev}
O.~Aharony, S.S.~Gubser, J.M.~Maldacena, H.~Ooguri and Y.~Oz,
\emph{Large-$N$ field theories, string theory and gravity},
\prep{323}{2000}{183} [\hepth{9905111}].

\bibitem{buch}
I.L.~Buchbinder, A.Y.~Petrov and A.A.~Tseytlin, \emph{Two-loop $N=4$
  super Yang-Mills effective action and interaction between
  d3-branes}, \npb{621}{2002}{179} [\hepth{0110173}].

\bibitem{beren}
D.~Berenstein, J.M.~Maldacena and H.~Nastase, \emph{Strings in flat
  space and $pp$ waves from $N=4$ super Yang-Mills},
\jhep{04}{2002}{013} [\hepth{0202021}].

\bibitem{mz}
J.A.~Minahan and K.~Zarembo, \emph{The Bethe-ansatz for $N=4$ super
  Yang-Mills}, \jhep{03}{2003}{013} [\hepth{0212208}].

\bibitem{beis}
N.~Beisert and M.~Staudacher, \emph{The $N=4$ SYM integrable super
  spin chain}, \npb{670}{2003}{439} [\hepth{0307042}].

\bibitem{beis2}
N. Beisert, C. Kristjansen, M. Staudacher, \emph{The dilatation operator of
conformal $N=4$ super Yang-Mills theory}, \npb{664}{2003}{131}
[\hepth{0303060}].

\bibitem{beis3}
N. Beisert, J.A. Minahan, M. Staudacher and K. Zarembo,
\emph{Stringing spins and spinning strings}, \jhep{09}{2003}{010}
[\hepth{0306139}].

\bibitem{beis4}
V.A.~Kazakov, A.~Marshakov, J.A.~Minahan and K.~Zarembo,
\emph{Classical/quantum integrability in AdS/CFT},
\jhep{05}{2004}{024} [\hepth{0402207}].

\bibitem{beis5}
N. Beisert, \emph{The dilatation operator of $N=4$ super Yang-Mills
theory and integrability}, \prep{405}{2005}{1} [\hepth{0407277}];
\emph{Higher-loop integrability in $N=4$ gauge theory},
\emph{Comptes Rendus Physique} {\bf 5} (2004) 1039 [\hepth{0409147}].


\bibitem{witt2}
E.~Witten, \emph{Perturbative gauge theory as a string theory in
  twistor space}, \cmp{252}{2004}{189} [\hepth{0312171}].

\bibitem{vol}
R.~Roiban and A.~Volovich, \emph{Yang-Mills correlation functions from
  integrable spin chains}, \jhep{09}{2004}{032} [\hepth{0407140}].

\bibitem{agarwal}
A.~Agarwal and S.G.~Rajeev, \emph{Yangian symmetries of matrix models
  and spin chains: the dilatation operator of $N=4$ SYM},
\hepth{0409180}.

\bibitem{sato}
M.~Sato and A.~Tsuchiya, \emph{Born-Infeld action from supergravity},
\ptp{109}{2003}{687} [\hepth{0211074}];
\emph{Hamilton-Jacobi method and effective actions of D-brane and
  m-brane in supergravity}, \npb{671}{2003}{293} [\hepth{0305090}];
\emph{A note on Hamilton-Jacobi formalism and D-brane effective
  actions}, \plb{579}{2004}{217} [\hepth{0310125}];
\emph{M5-brane effective action as an on-shell action in
  supergravity}, \jhep{11}{2004}{067} [\hepth{0410261}].

\bibitem{frot}
S.~Frolov and A.A.~Tseytlin, \emph{Multi-spin string solutions in
  $AdS_5\times S^5$}, \npb{668}{2003}{77} [\hepth{0304255}].

\bibitem{ts2}

A.A.~Tseytlin, \emph{Spinning strings and AdS/CFT duality},
\hepth{0311139}.

\bibitem{add1}
G. Arutyunov and M. Staudacher, \emph{Matching higher conserved
charges for strings and spins}, \jhep{03}{2004}{004}
[\hepth{0310182}];
\emph{Two-loop commuting charges and the string/gauge duality},
\hepth{0403077}.

\bibitem{add2}
G. Arutyunov, S. Frolov and M. Staudacher, \emph{Bethe ansatz for
quantum strings}, \jhep{10}{2004}{016} [\hepth{0406256}].

\bibitem{add3}
I.~Swanson, \emph{Quantum string integrability and AdS/CFT},
\hepth{0410282}.

\bibitem{add4}
G. Arutyunov and S. Frolov, \emph{Integrable hamiltonian for classical
strings on $AdS_5 \times S^5$}, \hepth{0411089}.

\bibitem{PBR}
I.~Bena, J.~Polchinski and R.~Roiban, \emph{Hidden symmetries of the
  $AdS_5\times S^5$ superstring}, \prd{69}{2004}{046002}
[\hepth{0305116}].

\bibitem{fubini}
S.~Fubini, A.J.~Hanson and R.~Jackiw, \emph{New approach to field
  theory}, \prd{7}{1973}{1732}.

\bibitem{nb}
N.~Beisert, S.~Frolov, M.~Staudacher and A.A.~Tseytlin,
\emph{Precision spectroscopy of AdS/CFT}, \jhep{10}{2003}{037}
     [\hepth{0308117}].

\bibitem{dnw}
L.~Dolan, C.R.~Nappi and E.~Witten, \emph{A relation between
  approaches to integrability in superconformal Yang-Mills theory},
\jhep{10}{2003}{017} [\hepth{0308089}];\\
\emph{Yangian symmetry in $D=4$ superconformal Yang-Mills theory},
\hepth{0401243}.

\bibitem{mackay}
N.J.~MacKay, \emph{Introduction to yangian symmetry in integrable
  field theory}, \hepth{0409183}.

\bibitem{msw}
G.~Mandal, N.V.~Suryanarayana and S.R.~Wadia, \emph{Aspects of
  semiclassical strings in $AdS_5$}, \plb{543}{2002}{81}
[\hepth{0206103}].

\bibitem{chiral}
E.~Br\'{e}zin, C.~Itzykson, J.~Zinn-Justin and J.B.~Zuber,
\emph{Remarks about the existence of nonlocal charges in two-
  dimensional models}, \plb{82}{1979}{442}.

\bibitem{MT}
R.R.~Metsaev and A.A.~Tseytlin, \emph{Type IIB superstring action in
  $AdS_5\times S^5$ background}, \npb{533}{1998}{109}
[\hepth{9805028}].

\bibitem{krr}
R.~Kallosh, J.~Rahmfeld and A.~Rajaraman, \emph{Near horizon
  superspace}, \jhep{09}{1998}{002} [\hepth{9805217}].

\bibitem{rs}
R.~Roiban and W.~Siegel, \emph{Superstrings on $AdS_5\times S^5$
  supertwistor space}, \jhep{11}{2000}{024} [\hepth{0010104}].

\bibitem{KEK}
M.~Hatsuda and K.~Yoshida, \emph{Classical integrability and super
  yangian of superstring on $AdS_5\times S^5$}, \hepth{0407044}.

\bibitem{nicolai}
V.~Belinskii and V.~Zakharov, \emph{Integration of the Einstein
equations by means of the inverse scattering problem technique and
construction of exact soliton solutions}, \newjournal{Sov.\ Phys.\
JETP}{SPHJA}{48}{1978}{985};\\
H.~Nicolai, \emph{Schladming lectures}, H.~Mitter and H.~Gausterer
eds., Springer-Verlag, Berlin 1991;\\
H.~Nicolai, D.~Korotkin and H.~Samtleben, \emph{Integrable classical
  and quantum gravity}, \hepth{9612065}.

\bibitem{BBHZZ}
N.~Berkovits, M.~Bershadsky, T.~Hauer, S.~Zhukov and B.~Zwiebach,
\emph{Superstring theory on $AdS_2\times S^2$ as a coset
  supermanifold}, \npb{567}{2000}{61} [\hepth{9907200}];\\
M.~Bershadsky, S.~Zhukov and A.~Vaintrob, \emph{${\rm PSL}(n|n)$ sigma model
  as a conformal field theory}, \npb{559}{1999}{205}
[\hepth{9902180}].

\bibitem{kr2}
R.~Kallosh and J.~Rahmfeld, \emph{The gs string action on $AdS_5\times
  S^5$}, \plb{443}{1998}{143} [\hepth{9808038}].

\bibitem{KalTs}
R.~Kallosh and A.A.~Tseytlin, \emph{Simplifying superstring action on
  $AdS_5\times S^5$}, \jhep{10}{1998}{016} [\hepth{9808088}].

\bibitem{kac3}
V.G.~Kac, \emph{A sketch of Lie superalgebra theory},
\cmp{53}{1977}{31}.

\bibitem{dewitt}
B.~DeWitt, \emph{Supermanifolds}, Cambridge University Press, 1992.

\bibitem{corn}
J.F.~Cornwell, \emph{Croup theory in physics}, vol.~3,
\emph{Supersymmetries and infinite-dimensional algebras}, Academic
Press, 1989.

\bibitem{jmj}
J.~Maharana and J.H.~Schwarz, \emph{Noncompact symmetries in string
  theory}, \npb{390}{1993}{3} [\hepth{9207016}].

\bibitem{DMM}
A.~Das, J.~Maharana, A.~Melikyan, \emph{Duality, monodromy and
  integrability of two dimensional string effective action},
\prd{65}{2002}{126001};
\emph{Duality and integrability of two dimensional string effective
  action}, \plb{533}{2002}{146};
\emph{Colliding string waves and duality}, \plb{518}{2002}{306}.

\bibitem{bm}
P.~Breitenlohner and D.~Maison, \emph{On the Geroch group},
\emph{Inst.\ H.\ Poincar\'{e}}, {\bf 46} (1987) 215;\\
F.J.~Ernst, A.~Garcia and I.~Hauser, \emph{Colliding gravitational
  plane waves with noncollinear polarization, 1},
\jmp{28}{1987}{2155}.

\bibitem{KS}
D.~Korotkin and H.~Samtleben, \emph{Yangian symmetry in integrable
  quantum gravity}, \npb{527}{1998}{657} [\hepth{9710210}].

\bibitem{HENTEI}
M.~Henneaux and C.~Teitelboim, \emph{Quantization of gauge systems},
Princeton University Press, 1994.

\bibitem{CZ}
T.L.~Curtright and C.K.~Zachos, \emph{Nonlocal currents for
  supersymmetric nonlinear models}, \prd{21}{1980}{411}.

\bibitem{dasbnm}
A.~Das, J.~Barcelos-Neto and J.~Maharana, \emph{Algebra of charges in
  the supersymmetric nonlinear $\sigma$ model},
\newjournal{Zeitschrift f\"ur Physik}{ZEPYA}{C30}{1986}{401}.

\bibitem{jm}
J.~Maharana, \emph{Infinite sequence of nonlocal conserved currents in
  a supersymmetric grassmannian model}, \lmp{8}{1984}{289};
\emph{The canonical structure of the supersymmetric nonlinear sigma model in
the constrained Hamiltonian formalism}, 
\emph{Ann.\ Inst.\ H.\ Poincar\'{e}} {\bf 45} (1986) 231.

\bibitem{FT}
L.D.~Faddeev, L.A.~Takhtajan, \emph{Hamiltonian methods in the
  theory of solitons}, Springer Verlag, Berlin, 1987;\\ 
L.~Faddeev, \emph{Integrable models in $(1+1)$-dimensional quantum
  field theory}, in \emph{Les Houches, lectures, recent advances in
  field theory and statistical mechanics}, Elsevier Science
Publishers, 1984.

\bibitem{das}
A.~Das, \emph{Integrable models}, World Scientific, Singapore, 1989.

\bibitem{pes1}
I.~Pesando, \emph{A kappa gauge fixed type-IIB superstring action on
  $AdS_5\times S^5$}, \jhep{11}{1998}{002} [\hepth{9808020}].

\bibitem{pes2}
I.~Pesando, \emph{On the fixing of the kappa gauge symmetry on AdS and
  flat background: the lightcone action for the type-IIB string on
  $AdS_5\times S^5$}, \plb{485}{2000}{246} [\hepth{9912284}].

\bibitem{Maillet}
J.M.~Maillet, \emph{New integrable canonical structures in
  two-dimensional models}, \npb{269}{1986}{54}.

\bibitem{NIDUN}
A.~Duncan, H.~Nicolai, M.~Niedermaier, \emph{On the Poisson bracket
  algebra of monodromy matrices}, DESY {\bf 89-100} (1989).

\bibitem{FARESH}
L.~Faddeev, N.~Reshetikhin, \emph{Integrability of the principal
  chiral field model in $(1+1)$ dimensions}, \ap{167}{1986}{227}.

\end{thebibliography}
\end{document}